\documentclass[sigconf]{acmart} 
\copyrightyear{2019} 
\acmYear{2019} 
\acmConference[RecSys '19]{Thirteenth ACM Conference on Recommender Systems}{September 16--20, 2019}{Copenhagen, Denmark}
\acmBooktitle{Thirteenth ACM Conference on Recommender Systems (RecSys '19), September 16--20, 2019, Copenhagen, Denmark}
\acmPrice{15.00}
\acmDOI{10.1145/3298689.3347011}
\acmISBN{978-1-4503-6243-6/19/09}

\usepackage{amssymb}
\usepackage{algorithm}
\usepackage{algorithmic}
\usepackage{bibentry}
\usepackage{amsfonts}
\usepackage{subfigure}
\usepackage{multirow}
\usepackage{amsmath}
\usepackage{graphicx}
\usepackage{epsfig}
\usepackage{color}
\usepackage{epstopdf}
\usepackage{rotating}
\usepackage{caption}
\usepackage{multicol}
\usepackage{amssymb}
\usepackage{graphicx}
\usepackage{bbding}
\usepackage{balance}

\begin{document}

\title{Deep Social Collaborative Filtering}

\author{Wenqi Fan}
\affiliation{%
  \institution{Department of Computer Science}
  \state{City University of Hong Kong}
}
\email{wenqifan03@gmail.com}

\author{Yao Ma}
\affiliation{%
  \institution{Data Science and Engineering Lab}
  \state{Michigan State University}
}
\email{mayao4@msu.edu}

\author{Dawei Yin}
\affiliation{%
  \institution{JD.com}
}
\email{yindawei@acm.org	}

\author{Jianping Wang}
\affiliation{%
  \institution{Department of Computer Science}
  \state{City University of Hong Kong}
}
\email{jianwang@cityu.edu.hk}

\author{Jiliang Tang}
\affiliation{%
  \institution{Data Science and Engineering Lab}
  \state{Michigan State University}
}
\email{tangjili@msu.edu}

\author{Qing Li}
\affiliation{%
  \institution{Department of Computing}
  \state{The Hong Kong Polytechnic University}
}
\email{csqli@comp.polyu.edu.hk}

\begin{abstract}
Recommender systems are crucial to alleviate the information overload problem in online worlds. Most of the modern recommender systems capture users' preference towards items via their interactions based on collaborative filtering techniques. In addition to the user-item interactions, social networks can also provide useful information to understand users' preference as suggested by the social theories such as homophily and influence. Recently, deep neural networks have been utilized for social recommendations, which facilitate both the user-item interactions and the social network information. However, most of these models cannot take full advantage of the social network information. They only use information from direct neighbors, but distant neighbors can also provide helpful information. Meanwhile, most of these models treat neighbors' information equally without considering the specific recommendations. However, for a specific recommendation case, the information relevant to the specific item would be helpful. Besides, most of these models do not explicitly capture the neighbor's opinions to items for social recommendations, while different opinions could affect the user differently. In this paper, to address the aforementioned challenges, we propose {\bf DSCF}, a {\bf D}eep {\bf S}ocial {\bf C}ollaborative {\bf F}iltering framework, which can  exploit the social relations with various aspects for recommender systems. Comprehensive experiments on two-real world datasets show the effectiveness of the proposed framework.

\end{abstract}


\maketitle

\section{Introduction}
Recommender systems play a crucial role to alleviate the information overload in the era of information explosion. Collaborative filtering is one of the most popular techniques to build modern recommender systems, which models users' preference towards items by utilizing the history of user-item interactions such as ratings~\cite{sarwar2001item}. In addition to the user-item interactions, social relations between users provide another stream of potential information of users' preference. As argued in social theories, people in social networks are influenced by their social connections, which leads to the homophily phenomenon of similar preference in social neighbors~\cite{bakshy2012role,o2002persuasion,fan2019deep,fan2019graph}. More specifically, information diffuses through social interactions and users tend to acquire and disseminate information through social networks. Thus, social relations can play an important role in describing the preferences of users, which, in turn, can help build good recommender systems. In fact, social relations have been shown to boost the performance of recommender systems~\cite{jamali2010matrix,tang2013social,fan2019graph,fan2019deep}.

Recent years have witnessed the great success of deep neural networks on various areas such as computer vision (CV)~\cite{wang2013learning}, speech recognition~\cite{hinton2012deep} and Natural Language Processing (NLP)~\cite{kalchbrenner2014convolutional}. It is not surprising that deep neural networks are adopted to enhance recommender systems. Some recent proposed recommender systems facilitate deep neural networks as feature learning tools to extract useful features from auxiliary information such as text description of items~\cite{wang2015collaborative,kim2016convolutional,Chen2018Neural}, audio of music~\cite{van2013deep,wang2014improving} and visual information of images~\cite{ZhaoLP016}, while others~\cite{He2017NCF} try to utilize deep neural networks to capture the non-linearity between user-item interactions. There are some recent works utilizing deep neural networks for social recommendations~\cite{deng2017deep,wang2017item,fan2019graph,fan2019deep,DeepSoR2018}. For example,  GraphRec~\cite{fan2019graph} proposes a graph neural networks framework for social recommendation, which aggregates both user-item interactions information and social interaction information when performing prediction; DASO~\cite{fan2019deep} harnesses the power of adversarial learning to dynamically generate "difficult" negative samples, learn the bidirectional mappings between the social domain and item domain.


Although the aforementioned deep social recommender systems facilitate the social network information to enhance the recommendation performance, they do not fully take advantage of social network information. First, most of them only involve direct neighbors, while information from users that are a few hops away could also be helpful~\cite{tang2013social, tang2013exploiting}. The reasons are as follows: 1) information is diffusing through the social network and users might be affected by indirect neighbors; and 2) users might refer to distant neighbors (or weak ties), when the direct neighbors cannot share useful information. Therefore, it is desired to consider the distant social relations for recommender systems. Second, most of the aforementioned methods treat neighbors' information equally for all recommendation cases. However, not all information from neighbors are useful when the recommender system is performing a specific recommendation. For example, when predicting whether a user will purchase an iPhone X, the interactions between his/her friends and iPhone X or other iPhone related items might be helpful while the interactions between his/her friends and Nike shoes might not be relevant. Therefore, it is necessary to filter information from neighbors. Finally, most of the deep social recommender systems do not consider the users' opinions towards items, which are usually expressed in the forms of reviews or ratings. It is obvious that bad and good opinions from a user's friends would affect the user's decision in tremendously different ways. Hence, it is desired to carefully consider the opinions of user-item interactions.

While it is of great potential to sufficiently exploit the social network information for recommendations, it faces tremendous challenges. First, the social interactions in distant social relations are complex and it is difficult to properly extract helpful information for recommendations. Second, it is not trivial to select relevant information from neighbors, as they could have interactions with many different items. Finally, it is challenging to capture the user's opinions while modeling the user-item interactions. In this paper, to tackle the aforementioned challenges, we propose a deep social collaborative filtering framework DSCF, which can sufficiently exploit the social network information for recommendations. Our contributions can be summarized as follows:
\begin{itemize}
\item We propose a principle way based on deep neural networks to extract helpful information from distant social relations for recommendations;
\item We introduce a novel way to capture user's opinions while modeling user-item interactions;
\item We propose a deep social collaborative filtering framework which can sufficiently exploit social network information for recommendations; and
\item We conduct comprehensive experiments on two real-world datasets to show the effectiveness of the proposed framework.
\end{itemize}

The remainder of this paper is organized a follows. We introduce the proposed framework in Section~\ref{sec: framework}. In Section~\ref{sec:Experiments}, we conduct experiments on two real-work datasets to illustrate the effectiveness of the proposed method. In Section~\ref{sec:relatedwork}, we review work related to our framework. Finally, we conclude our work with future directions in Section~\ref{sec:conclusion}.

\section{The proposed framework}
\label{sec: framework}
\begin{figure}[htbp]
\centering
{\includegraphics[width=0.99\linewidth]{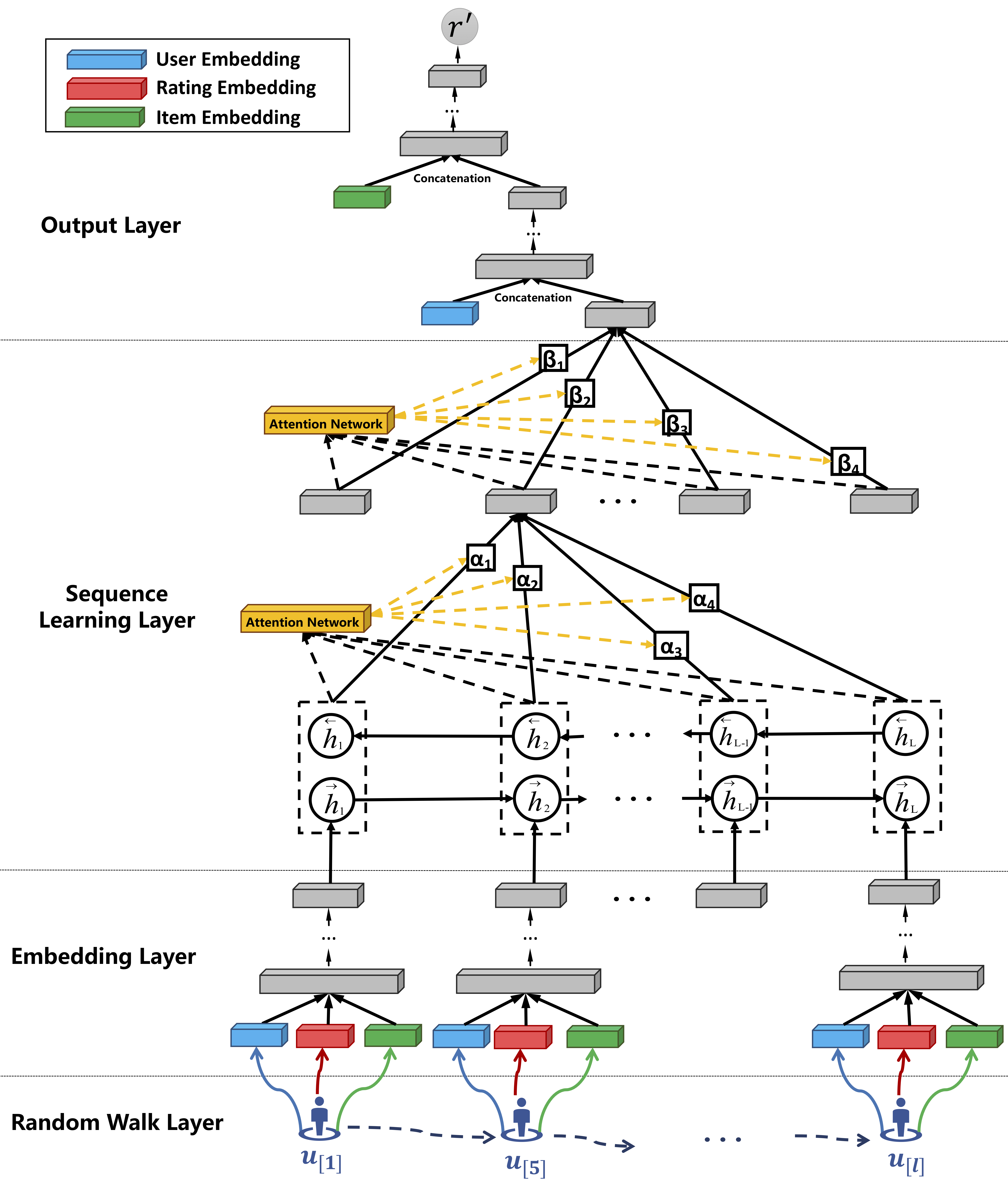}}
\caption{An overview of the proposed framework.}\label{fig:overview}
\end{figure}

In this section, we introduce the proposed deep social collaborative filtering framework {\bf DSCF}. As discussed earlier, to exploit social networks for recommendations, we need to (a) consider information from not only direct neighbors but also distant neighbors; (b) select relevant information of each neighbor for recommending a specific item; and (c) capture neighbor's opinions towards items when modeling user-item interactions. An overview of the proposed framework is demonstrated in Figure~\ref{fig:overview}. It consists of four layers -- the random walk layer that is designed for addressing challenges (a) and (b), the embedding layer that is designed for solving the challenge (c), the sequence learning layer and the output layer. Next we will give details of each layer.

Before introducing the details of each layer, we first introduce definitions and notations that are used through the paper. Let $ \mathcal{U}= \left \{ u_1, u_2, ..., u_N \right \}$ and $ \mathcal{V} = \left \{ v_1, v_2, ..., v_M \right \}$ denote the sets of users and items respectively, where $N$ is the number of users, and $M$ is the number of items. Let ${\bf R} \in  \mathbb{R}^{N\times M}$ be the rating matrix (or the user-item interaction matrix), where the $i,j$-th element $r_{i,j}$ is the rating score of item $v_j$ given by user $u_i$. If the user $u_i$ has not rated the item $v_j$, then $ r_{i,j}$ is set to $0$, which means the rating is unknown. The social network between users can be described by a matrix $ {\bf T}\in \mathbb{R}^{N\times N}$, where ${\bf T}_{i,j}=1$ if there is a social relation between user $u_i$ and user $u_j$, otherwise $0$. Given the rating matrix ${\bf R}$ and the social network ${\bf T}$, we aim to predict the unknown ratings in ${\bf R}$. As in the traditional collaborative filtering methods, we embed users and items to low-dimensional latent vectors. The embeddings for user $u_i$ and item $v_j$ are denoted as ${\bf p}_{[i]} \in \mathbb{R}^d$, ${\bf q}_{[j]}\in \mathbb{R}^d$ respectively, where $d$ is the length of the embedding.  

\subsection{The Random walk layer: generating item-aware social sequences}
In social recommendation, when we try to perform recommendation for a given user $u$, not only his/her direct neighbors can provide useful information, but also his/her distant neighbors that are within a few hops (or neighbors in his/her local neighborhood) can help. Furthermore, neighbors with different distance to the user $u$ are likely to be of different importance for the recommendation. Thus, it is also necessary to differentiate neighbors of $u$ according to their distance to user $u$ when including them for recommendations. Random walk is a popular tool to explore the local neighborhood of networks~\cite{tadic2003exploring,lovasz1993random}. Additionally, random walk explores the neighborhood in the form of node sequences (user sequences)~\cite{grover2016node2vec, perozzi2014deepwalk}, which naturally maintains the order of neighbors according to the distance to the user $u$. Thus, we can effectively utilize random walk to generate distant user sequences from social networks. More specifically, the user sequence can be generated by a random walk starting from user $u$ and ending after $l$ steps, where $l$ is the length of the random walk. The generated user sequence can be denoted as $S^u_{(i)} = \{u_{[1]},\dots, u_{[l]}\}$, where the subscript $(i)$ indicates ${S^u_{(i)}}$ is the $i$-th user sequence generated for user $u$ as we need to generate multiple user sequences to sufficiently explore the neighborhood of $u$ and $[k]$ means that the user $u_{[k]}$ is the $k$-th user in the user sequence.

While the user sequences contain the information of neighbors, they are not specified for a given recommendation case, i.e., predicting preference of user $u$ on the item $v$, as such information is shared by all the recommendation cases involving user $u$. However, not all information from the neighbors is helpful for recommending the specific item $v$. Only that information related to this item $v$ would be useful. Thus, we need to select an item related to item $v$ for each user in the generated user sequences and form an item-aware social sequence, denoted as $S^{u,v}_{(i)} = \{(u_{[1]}, v_{[1]}),\dots, (u_{[l]},v_{[l]})\}$.  Note that only the most relevant item is exploited for a specific recommendation case. The reasons are two-fold. First, the most relevant item is most important to affect the decision making of a target item (item $v$), while other items may not be helpful since they may bring in noise. Second, multiple user sequences are generated by the random walk process to sufficiently explore different relevant items for a specific recommendation case, which, in turn, can help form these item-aware social sequences. More specifically, for each user $u_{[k]}$ in one user sequence, we choose the item $v_{[k]}$ from the set of items that have been interacted with user $u_{[k]}$ as:
\begin{align}
v_{[k]} = argmax_{v_h\in {\mathcal{V}}_{u_{[k]}}}
sim(v_h,v),
\end{align}
where $\mathcal{V}_{u_{[k]}}$ denotes the set of items interacted with user $u_{[k]}$ and $sim(v_h,v)$ is a function to measure the similarity between item $v_h$ and item $v$. In this paper, we empirically select cosine similarity as follows
\begin{eqnarray}
sim(v_h,v) &=& \frac{{\bf x}_h^T{\bf x}_{v}}{|{\bf x}_h||{\bf x}_{v}|}, \\
{\bf x}_{m} &=& f(v_m).
\end{eqnarray}
where $f$ function is to generate appropriate features ${\bf x_m}$ for item $v_m$. Different features sources, such as the textual descriptions, the visual content of images and the user-item interactions, could be used to represent the items. In this paper, we adopt the user-item interactions to represent the items since the auxiliary information such as textual descriptions and visual content is not available. More specifically, we use the item embeddings learned by NeuMF~\cite{He2017NCF} as the item features to measure similarity between items.

The set of all item-aware social sequences generated for predicting the rating of ${(u,v)}$ is $\mathcal{S}^{u,v} = \{S^{u,v}_{(i)}\}_{i=1}^H$, where $H$ is the number of social sequences generated for this recommendation case.

\begin{figure}[htbp]
\centering
{\includegraphics[width=0.99\linewidth]{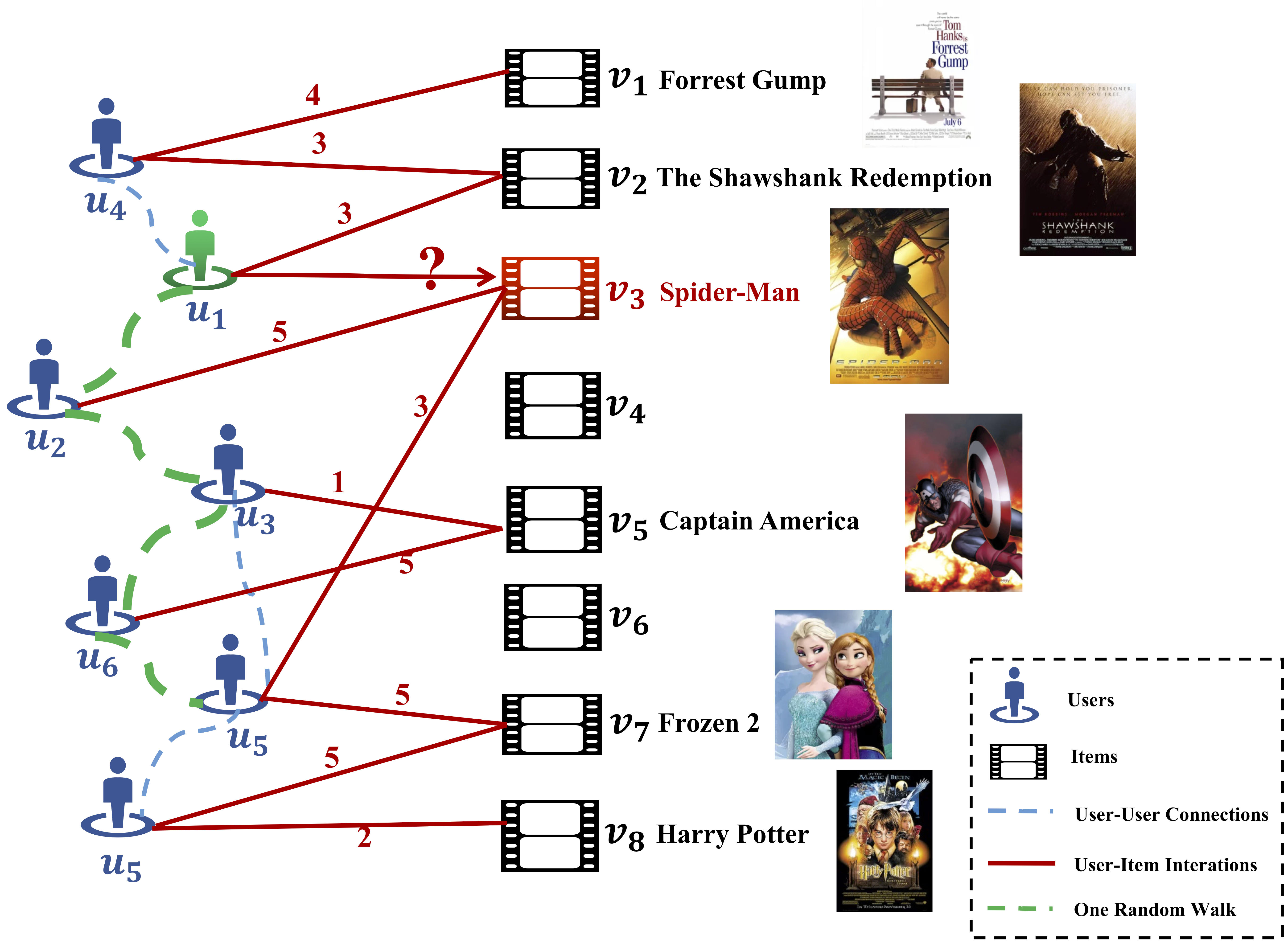}}
\caption{An illustration example of generating item-aware social sequences. Note that the number on the edges of user-item interactions denotes the opinions (or rating score) of users on the items via the interactions. There are 5 different rating levels.} \label{fig:item-aware-social}
\end{figure}

The advantages of the item-aware social sequences for predicting interaction between users $u$ and items $v$ are twofolds. First, the social sequences contain not only direct neighbors but also distant neighbors. Second, these sequences are specific for the recommendation from $u$ to $v$. An illustration example of the process of generating item-aware social sequence is shown in Figure~\ref{fig:item-aware-social}. We are predicting the rating of user $u_1$ to item $v_3$ (Spider-man). As shown in figure, starting from source user $u_1$, we perform our random walk on the direct neighbors. The random walk is employed to generate possible user sequence, denoted as $S^{u_1}_{(1)}=\left \{ u_{[2]}, u_{[3]}, u_{[6]}, u_{[7]} \right \}$. For each user in the user sequence, we need to collect the most similar item to $v_3$. The generated item-aware social sequence is $S^{u_1,v_3}_{(1)} = \left \{(u_{[2]},v_{[3]}), (u_{[3]},v_{[5]}),(u_{[6]}, v_{[5]}),(u_{[7]}, v_{[3]}) \right \}$. To prevent clutter, here, we suppose that item $v_5$ (Captain America) is the most similar to the item $v_3$ (Spider-Man) in our example, and the length of random walk is $4$.

\subsection{The embedding layer: modeling user-item interactions}

The item-aware sequences consist of user-item interactions from the user's neighbors, hence, we need to first model the user-item interactions. When modeling the user-item interactions, it is important to carefully consider the opinions the users expressed on the interactions. Obviously, bad and good opinions from the user's social neighbors can affect the user's opinion towards the item in tremendously different ways. Thus, we propose to include the user's opinion towards the item when modeling the user-item interaction in the sequence. The opinions are usually expressed in the form of ratings. For example, as shown in Figure~\ref{fig:item-aware-social},  both user $u_3$ and user $u_6$ interact with the same item $v_5$ (Captain America); however, user $u_6$ likes $v_5$ while user $u_3$ dislikes $v_5$.

To model the ratings, we propose to embed each discrete rating value into a rating embedding vector. Therefore, if there are $I$ different rating levels, there would be $I$ rating embedding vectors. Note that the rating embeddings are also parameters of the framework. The rating embedding of the rating value $o$ is denoted as ${\bf r}_{\{o\}} \in \mathbb{R}^d$, with $d$ the embedding length. For an interaction $(u_{[k]},v_{[k]})$ in the item-aware social sequence $S^{u,v}_{(i)}$, the non-zero rating score of this interaction can be found in the rating matrix ${\bf R}$ and let us denote it as $o_{u_{[k]},v_{[k]}}$. Then the corresponding rating embedding is ${\bf r}_{\{o_{u_{[k]},v_{[k]}}\}}$, which we denote as ${\bf r}_{[k]}$ for convenience. The interaction between user and item is highly non-linear, and including the rating information further adds the complexity. Hence, we use a multi-layer perception (MLP) to fuse the interaction information with the rating information. The MLP takes the concatenation of user embedding ${\bf p}_{[k]}$, rating embedding ${\bf r}_{[k]}$, item embedding ${\bf q}_{[k]}$ as input and output the user-item interaction embedding ${\bf e}_{[k]}$ of interaction $({u_{[k]}, v_{[k]}})$.
The procedure can be briefly represented as follows
\begin{align}
{\bf e}_{[k]} = g_{u,r,v}([{\bf p}_{[k]}, {\bf r}_{[k]}, {\bf q}_{[k]}])
\end{align}
where $[{\bf p}_{[k]}, {\bf r}_{[k]}, {\bf q}_{[k]}]$ denotes the concatenation of ${\bf p}_{[k]}, {\bf r}_{[k]}, {\bf q}_{[k]}$.


Following this procedure, we process each sequence $S^{u,v}_{(i)} = \{(u_{[1]},v_{[1]}),\dots, (u_{[l]},v_{[l]})\}$ and get a sequence of fused interaction embedding $E^{u,v}_{(i)} = \{{\bf e}_{[1]},\dots, {\bf e}_{[l]}\}$. The set of all sequences of fused interaction embedding from neighbors for predicting the rating of $(u,v)$ can be denoted as $\mathcal{E}^{u,v}$.


\subsection{The sequence learning layer: learning representation for item-aware social sequences}

After generating the item-aware social sequences for $(u,v)$ and transforming each user-item interaction with opinions information in the sequences to fused interaction embedding, we proceed to the sequence learning layer. The sequence learning layer aims to extract features for each sequence and then combine the extracted features of all the sequences to obtain a unified representation, which can be used to predict the rating for $(u,v)$ in the output layer.

As all the neighbors in the sequence would affect the prediction of $(u,v)$, for distant neighbors, we need to capture the distant social information between them and the user $u$. Furthermore, in social networks, users influence each other. Hence, we need to capture the bi-directional influence in the model. Recently, a bi-directional long short-term memory network (Bi-LSTM) based language model~\cite{yang2016hierarchical,Bahdanau2015Neural} has been proposed to capture the long-range bi-directional semantic dependencies between words in sentence in NLP domain. Inspired by these model, we regard the sequence as a ``sentence'' and elements in this sequence as ``words'' and adopt a similar Bi-LSTM model to extract features from the sequence of fused interaction embeddings. The bi-directional LSTM contains the forward LSTM $\overrightarrow{LSTM}$ which reads the sequence $E^{u,v}_{(i)}$ from $\mathbf{e}_{[1]}$ to $\mathbf{e}_{[l]}$, and a backward LSTM $\overleftarrow{LSTM}$ which reads from $\mathbf{e}_{[l]}$ to $\mathbf{e}_{[1]}$,
\begin{eqnarray}
\label{eq:BiLSTM}
\overrightarrow{{\bf h}^{(i)}_{[k]}} = \overrightarrow{LSTM}(E^{u,v}_{(i)}), k\in [1, l], \\
\overleftarrow{{\bf h}^{(i)}_{[k]}} = \overleftarrow{LSTM}(E^{u,v}_{(i)}), k\in [l, 1].
\end{eqnarray}
where $\overrightarrow{{\bf h}^{(i)}_{[k]}}$ and $\overleftarrow{{\bf h}^{(i)}_{[k]}}$ are hidden states of $\overrightarrow{LSTM}$, $\overleftarrow{LSTM}$, respectively.

These hidden states, which are corresponding to the neighbors in the sequence, are then combined using an attention mechanism~\cite{vaswani2017attention,fan2019graph,NIPS2017_7181} to generate the features ${\bf s}^{u,v}_{(i)}$ of the sequence $E^{u,v}_{(i)}$.
\begin{align}
{\bf s}^{u,v}_{(i)}  =\sum\limits_{k=1}^l \alpha_{k} {\bf h}^{(i)}_{[k]}.
\end{align}
where ${\bf h}^{(i)}_{[k]}$ is $[\overrightarrow{{\bf h}^{(i)}_{[k]}} \oplus \overleftarrow{{\bf h}^{(i)}_{[k]}}]$, the concatenation of $\overrightarrow{{\bf h}^{(i)}_{[k]}}$ and $\overleftarrow{{\bf h}^{(i)}_{[k]}}$.
Specially, we parameterize the attention weight $\alpha _{k}$ with one-layer network, and extract these user (neighbor)-item interaction embeddings that are important to learn representation for the item-aware social sequence. The normalized importance weight $\alpha_k$ is calculated through a Softmax function follows

\begin{eqnarray}
{\bf a }_{k} &=& tanh({\bf W}_a  \cdot  {\bf h}_{[k]} + {\bf b}_a),\\
\alpha _{k} &=& \frac{exp({\bf a}_{k}^{T}{\bf a}_u)}{\sum_j exp({\bf a}_{j}^{T}{\bf a}_u)}\label{eq:att_alpha}.
\end{eqnarray}
where the neighbor-level context vector ${\bf a}_u$ can be seen as a high level representation of a fixed query ``what is the informative neighbor-item interaction embedding?'' over all the neighbor-item interaction embeddings in the item-aware social  sequence. Note that the neighbor-level context vector ${\bf a}_u$ is parameters in the framework and needs to be jointly learned during the training process.


We then combine the representations of all the user-item interaction embedding sequences to generate the unified representation of item-aware social sequences for $(u,v)$ as
\begin{align}
{\bf s}^{u,v} = \sum\limits_{i=1}^H \beta_i   {\bf s}^{u,v}_{(i)}
\end{align}
where we adopt an attention mechanism to differentiate the importance weight $\beta _{i}$ of item-aware social sequences as follows
\begin{eqnarray}
{\bf z }_{i} &=& tanh({\bf W}_z  \cdot   {\bf s}^{u,v}_{(i)} + {\bf b}_z),\\
\beta _{i} &=& \frac{exp({\bf z}_{i}^{T}{\bf z}_u)}{\sum_j exp({\bf z}_{j}^{T}{\bf z}_u)}.
\end{eqnarray}
Similar to eq.~\eqref{eq:att_alpha}, ${\bf z}_u$ can be seen as a high level representation query ``which is the informative item-aware social sequence?'' over all the social sequences. 
The reason why we introduce two attentions is that not all user-item interaction with opinions information in one item-aware social sequence contribute equally to the representation of this item-aware social sequence; and not all these sequences contribute equally to the unified representation of item-aware social sequences for $(u,v)$.

\subsection{The output layer: rating prediction}

In the output layer, we will design recommendation tasks to learn model parameters. There are various recommendation tasks such as item ranking and rating predation. In this work, we apply the proposed DSCF model for the recommendation task of rating prediction. We finally make the prediction of rating score of the user $u$ to item $v$. The input of the output layer includes the user embedding ${\bf p}_{[u]}$, the item embedding ${\bf q}_{[v]}$ and the unified item-aware social representations ${\bf s}^{u,v}$ learned in the sequence learning layer. As shown in the output layer in Figure~\ref{fig:overview}, a multi-layer perception (MLP) is first used to combine the user embedding ${\bf p}_{[u]}$ and the unified item-aware social representations ${\bf s}^{u,v}$. Let us denote this MLP as $f_{u,s}$. Then, another MLP, which can be denoted as $f_{u,v}$, is used to predict the rating score of $(u,v)$. The prediction procedure, which takes ${\bf p}_{[u]}, {\bf q}_{[v]}, {\bf s}^{u,v}$ as input, can be represented as
\begin{align}
r'_{u,v} = f_{u,v}([{\bf q}_{[v]}, f_{u,s}([{\bf p}_{[u]},{\bf s}^{u,v}])]),
\end{align}
where $[,]$ denotes the concatenation operation, and $r'_{u,v}$ is the predicted rating from user $u$ to item $v$.



\subsection{Model Training}

To estimate parameters of the framework {\bf DSCF}, we need to specify an  objective function to optimize. Since the task we focus on in this work is rating prediction, a commonly used objective function is formulated as,
\begin{align}
Loss = \frac{1}{ 2 \left | \mathcal{O} \right | }  \sum _{(u_i,v_j) \in \mathcal{O}} (r'_{i,j} - r_{i,j})^2
\end{align}
\noindent where $\mathcal{O}$ denotes all the observed user-item interactions, $|\mathcal{O}|$ is the number of interactions in $\mathcal{O}$, and $r'_{i,j}$ is the predicted rating while $ r_{i,j}$ is the ground truth rating assigned by the user $u_i$ on the item $v_j$.

To optimize the objective function, we adopt the Adaptive Moment Estimation (Adam)~\cite{kingma2014adam} as the optimizer in our implementation. We also adopt the dropout  strategy ~\cite{srivastava2014dropout} to alleviate the overfitting issue in optimizing deep neural network models.

There are three embedding in our model, including item embedding $\mathbf{q}_j$, user embedding $\mathbf{p}_i$, and rating embedding $\mathbf{r}_o$. They are randomly initialized and jointly learned during the training stage.  We do not use one-hot vectors to represent each user and item, since the raw features are very large and highly sparse. By embedding high-dimensional sparse features into a low-dimensional latent space, the model can be easy to train~\cite{He2017NCF, wang2017irgan}. Rating embedding matrix $\mathbf{r}$ depends on the rating scale of the system. For example, for a 5-star rating system,  rating embedding matrix $\mathbf{r}$ contains 5 different embedding vectors to denote scores in $\left \{ 1,2,3,4,5 \right \}$.

\section{Experiments}
\label{sec:Experiments}
In this section, we conduct experiments to verify the effectiveness of our model. We first introduce the experimental settings, then discuss the results of the performance comparison of various recommender systems, and finally study the impact of different components in our model.



\subsection{Experimental Settings}
\subsubsection{Datasets}
In our experiments, two representative datasets Ciao and Epinions\footnote{ Both Ciao and Epinions datasets are available at: https://www.cse.msu.edu/$\sim$tangjili/trust.html} are utilized to verify the effectiveness of our model. They are taken from the product review sites Ciao (www.ciao.co.uk) and Epinions (www.epinions.com). Each site allows users to rate items,  and add friends to their `Circle of Trust'. Therefore, they provide a large amount of rating information and social information. The rating scale is from 1 to 5. We randomly initialize rating embedding with 5 different embedding vectors based on 5 scores in $\left \{ 1,2,3,4,5 \right \}$.  The statistics of these two datasets are presented in~\tablename ~\ref{tab:dataset}.

\begin{table}[htbp]
\centering
\caption{Statistics of the datasets}
\label{tab:dataset}
\begin{tabular}{l|c|c}
\hline
Dataset               & Ciao  &Epinions \\ \hline
\# of Users           & 7,317   &18,088 \\ \hline
\# of Items           & 104,975  &261,649 \\ \hline
\# of Ratings          & 283,319  &764,352\\ \hline
\# of Density(Ratings)          & 0.0368\% &0.0161\%  \\ \hline \hline
\# of Social Connections & 111,781 &355,813 \\ \hline
\# of Density(Social Relations) & 0.2087\% &0.1087\% \\ \hline
\end{tabular}
\end{table}

\subsubsection{Evaluation Metrics}
In order to evaluate the quality of the recommendation algorithms, two popular metrics are adopted to evaluate the predictive accuracy, namely Mean Absolute Error (MAE) and Root Mean Square Error (RMSE)~\cite{fan2019graph}. Smaller values of MAE and RMSE indicate better predictive accuracy. Note that small improvement in RMSE or MAE terms can have a significant impact on the quality of the top-few recommendations~\cite{koren2008factorization}.

\subsubsection{Baselines}

To evaluate the performance, we compared DSCF with three groups of methods including traditional recommender systems, traditional social recommender systems, and deep neural network based recommender systems. For each group, we select representative baselines and below we detail them.

\begin{itemize}
  \item \textbf{PMF}~\cite{salakhutdinov2007probabilistic}: \textbf{P}robabilistic \textbf{M}atrix \textbf{F}actorization utilizes user-item rating matrix only and models latent factors of users and items by Gaussian distributions.
  \item \textbf{SoRec}~\cite{ma2008sorec}: \textbf{So}cial \textbf{Rec}ommendation performs co-factorization on the user-item rating matrix and user-user social relations matrix.
  \item \textbf{SoReg}~\cite{ma2011recommender}: \textbf{So}cial \textbf{Reg}ularization models social network information as regularization terms to constrain the matrix factorization framework.
  \item \textbf{SocialMF}~\cite{jamali2010matrix}: It considers the trust information and propagation of trust information into the matrix factorization model for recommender systems.
  \item \textbf{TrustMF}~\cite{yang2017social}: This method adopts matrix factorization technique that maps users into two low-dimensional spaces: truster space and trustee space, by factorizing trust networks according to the directional property of trust.
  \item \textbf{NeuMF}~\cite{He2017NCF}: This method is a state-of-the-art matrix factorization model with neural network architecture. The original implementation is for recommendation ranking task and we adjust its loss to the squared loss for rating prediction.
  \item \textbf{DeepSoR}~\cite{DeepSoR2018}: This model employs deep learning to learn representations of each user from social relations, and to integrate them into probabilistic matrix factorization for rating prediction.
  \item  \textbf{GCMC+SN}~\cite{berg2017graph}: This model is a state-of-the-art recommender system with graph neural network architecture. In order to incorporate social network information into GCMC, we utilize the $node2vec$~\cite{grover2016node2vec} to generate user embedding as user side information, instead of using the raw feature social connections ($\mathrm{T} \in \mathbb{R} ^{n \times n}$) directly. The reason is that the raw feature input vectors is highly sparse and high-dimensional. Using the network embedding techniques can help compress the raw input feature vector to a low-dimensional and dense vector, then the model can be easy to train.
\end{itemize}

PMF and NeuMF are pure collaborative filtering model without social information for rating prediction, while the others are social recommendations. Besides, we compared DSCF with two state-of-the-art neural network based social recommender systems, i.e., DeepSoR, and CGMC+SN. 

\begin{table*}[htbp]
\centering
\caption{Performance comparison of different recommender systems}
\label{tab:baselines_results}
\begin{tabular}{|c|c|c|c|c|c|c|c|c|c|c|}
\hline
\multirow{2}{*}{Training}                                                   & \multirow{2}{*}{Metrics} & \multicolumn{9}{c|}{Algorithms}                                                        \\
\cline{3-11}
                                                                            &                          & PMF    & SoRec  & SoReg  & SocialMF & TrustMF & NeuMF  & DeepSoR & GCMC+SN  & DSCF \\
\hline
\multirow{2}{*}{\begin{tabular}[c]{@{}c@{}}Ciao\\ (60\%) \end{tabular}}      & MAE                      & 0.952  & 0.8489 & 0.8987 & 0.8353   & 0.7681  & 0.8251 & 0.7813  & 0.7697    &0.7501  \\
\cline{2-11}
                                                                            & RMSE                     & 1.1967 & 1.0738 & 1.0947 & 1.0592   & 1.0543  & 1.0824 & 1.0437  & 1.0221   & 1.0157 \\
\hline
\multirow{2}{*}{\begin{tabular}[c]{@{}c@{}}Ciao\\ (80\%) \end{tabular}}      & MAE                      & 0.9021 & 0.8410 & 0.8611 & 0.8270   & 0.7690  & 0.8062 & 0.7739  & 0.7526    & 0.7270 \\
\cline{2-11}
                                                                            & RMSE                     & 1.1238 & 1.0652 & 1.0848 & 1.0501   & 1.0479  & 1.0617 & 1.0316  & 0.9931   & 0.9867 \\
\hline\hline
\multirow{2}{*}{\begin{tabular}[c]{@{}c@{}} Epinions\\ (60\%) \end{tabular}} & MAE                      & 1.0211 & 0.9086 & 0.9412 & 0.8965   & 0.8550  & 0.9097 & 0.8520  & 0.8602   & 0.8427 \\
\cline{2-11}
                                                                            & RMSE                     & 1.2739 & 1.1563 & 1.1936 & 1.1410   & 1.1505  & 1.1645 & 1.1135  & 1.1004   & 1.0999 \\
\hline
\multirow{2}{*}{\begin{tabular}[c]{@{}c@{}}Epinions\\ (80\%) \end{tabular}}  & MAE                      & 0.9952 & 0.8961 & 0.9119 & 0.8837   & 0.8410  & 0.9072 & 0.8383  & 0.8590    & 0.8275 \\
\cline{2-11}
                                                                            & RMSE                     & 1.2128 & 1.1437 & 1.1703 & 1.1328   & 1.1395  & 1.1476 & 1.0972  & 1.0711   & 1.0667  \\
\hline
\end{tabular}
\end{table*}


\subsubsection{Parameter Settings}
We implemented our proposed model in Pytorch\footnote{https://pytorch.org/}.
For each dataset, we used $x$\% as a training set to learning parameters, $(1-x\%)/2$ as a validation set to tune hyper-parameters, and $(1-x\%)/2$ as a testing set for the final performance comparison, where $x$ was varied as $\left \{ 80\%, 60\% \right \}$~\cite{fan2019graph}. For the embedding size $d$, we tested the value of $\left \{ 8, 16, 32, 64, 128, 256 \right \}$. The batch size and learning rate were searched in $\left \{ 16, 32, 64, 128, 512 \right \} $ and $\left \{ 0.0005, 0.001, 0.005, 0.01, 0.05, 0.1 \right \}$, respectively. Moreover, we empirically set the size of the hidden layer the same as the embedding size (the dimension of the latent factor) and the activation function as ReLU. Without special mention, we employed three hidden layers for all the neural components. The early stopping strategy was performed, where we stopped training if the RMSE on validation set increased for 5 successive
epochs. The parameters for the baseline algorithms were initialized as suggested in the corresponding papers, and were then carefully tuned to achieve optimal performance.

\subsection{Performance Comparison}
We first compare the recommendation performance of all methods. Table~\ref{tab:baselines_results} shows the overall rating prediction error $w.r.t.$ RMSE and MAE among the recommendation methods on Ciao and Epinions datasets, respectively. We have the following findings:

\begin{itemize}
  \item  SoRec, SoReg, SocialMF and TrustMF improve over PMF. All of these methods are based on matrix factorization. SoRec, SoReg, SocialMF and TrustMF leverage both the user-item interactions and social information; while PMF only utilizes user-item interactions. These improvements show the effectiveness of incorporating social information for recommender systems.
   \item NeuMF achieves much better performance than PMF. Both of them utilize the user-item interactions only. NeuMF is based on deep architecture; while PMF is a traditional method with shallow architecture. This suggests the power of employing deep architecture on the task of recommendation.
   \item  Two deep models, DeepSoR and GCMC+SN, obtain better performance than SoRec, SoReg, SocialMF, and TrustMF, which are based on matrix factorization with shallow architecture. These improvements further reflect the power of employing deep architecture on the task of recommendation.

\item DSCF outperforms NeurMF. This result further supports that social information is complementary to user-item interactions for recommendation.

   \item Our model DSCF consistently outperforms all the baseline methods. Compared with DeepSoR and GCMC+SN, our model proposes advanced model components to integrate user-item interactions and social information. In addition, our model introduces ways to capture user's opinions while modeling user-item interactions. We will provide further investigations to better understand the contributions of model components to the proposed framework in the following subsection.

\end{itemize}


\begin{figure}[t]
\centering
{\subfigure[RMSE]
{\includegraphics[width=0.45\linewidth]{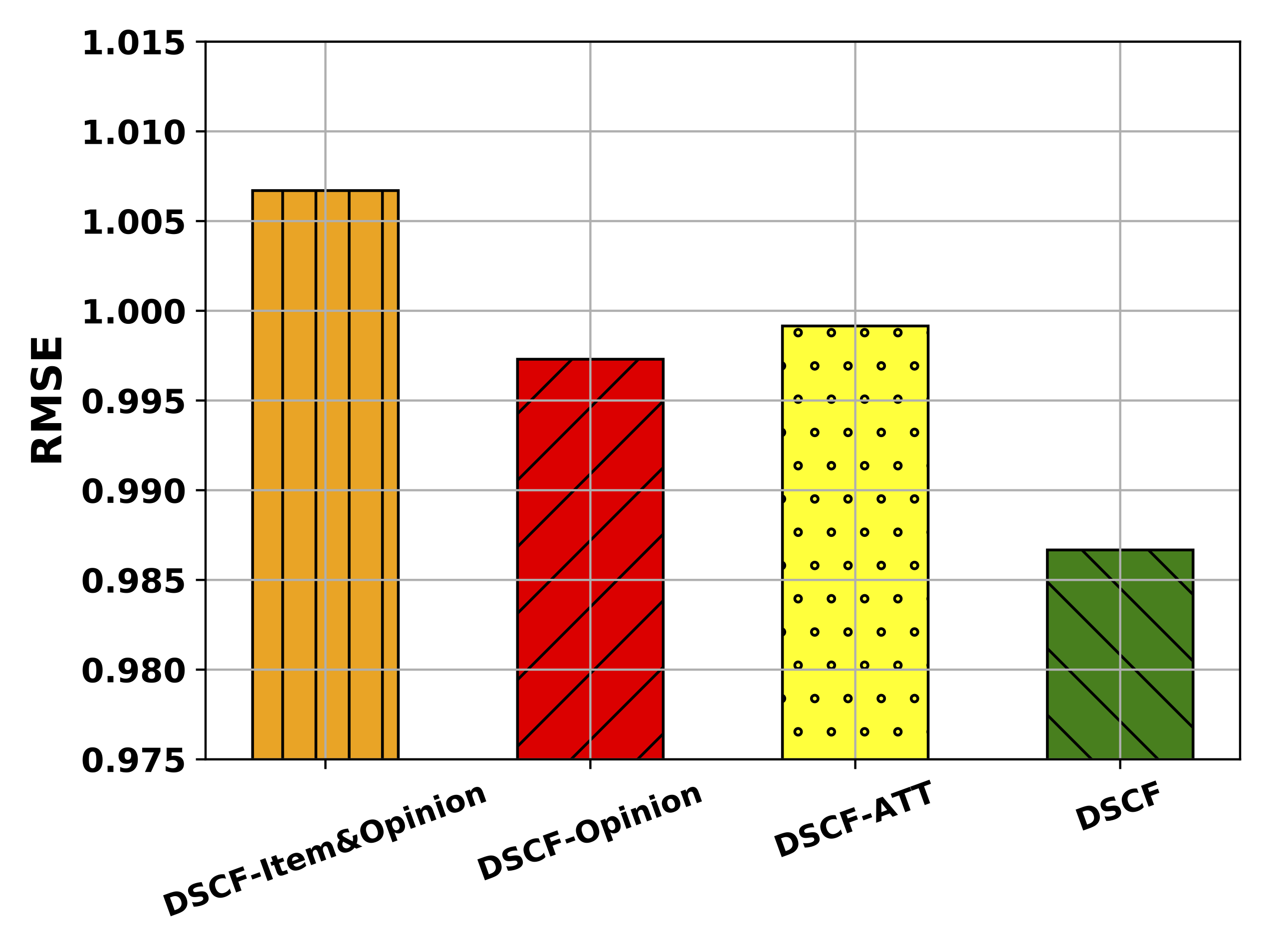}\label{fig:ciao_uvr_rmse}}}
{\subfigure[MAE]
{\includegraphics[width=0.45\linewidth]{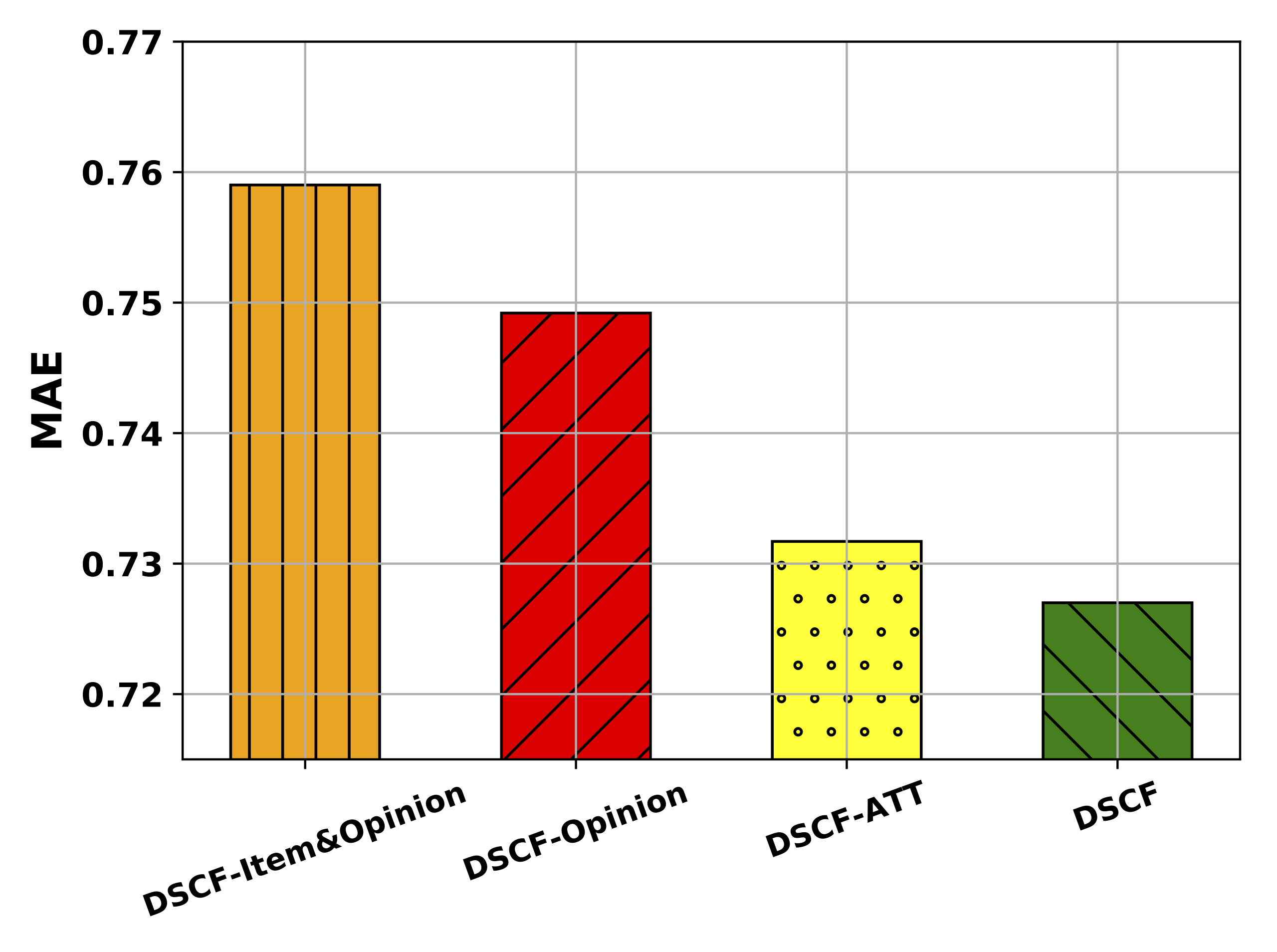}\label{fig:ciao_uvr_mae}}}
\caption{Component Analysis on Ciao dataset. DSCF-* means the component * is removed in DSCF.}\label{fig:ciao_uvr}
\end{figure}
\begin{figure}[t]
\centering
{\subfigure[RMSE]
{\includegraphics[width=0.45\linewidth]{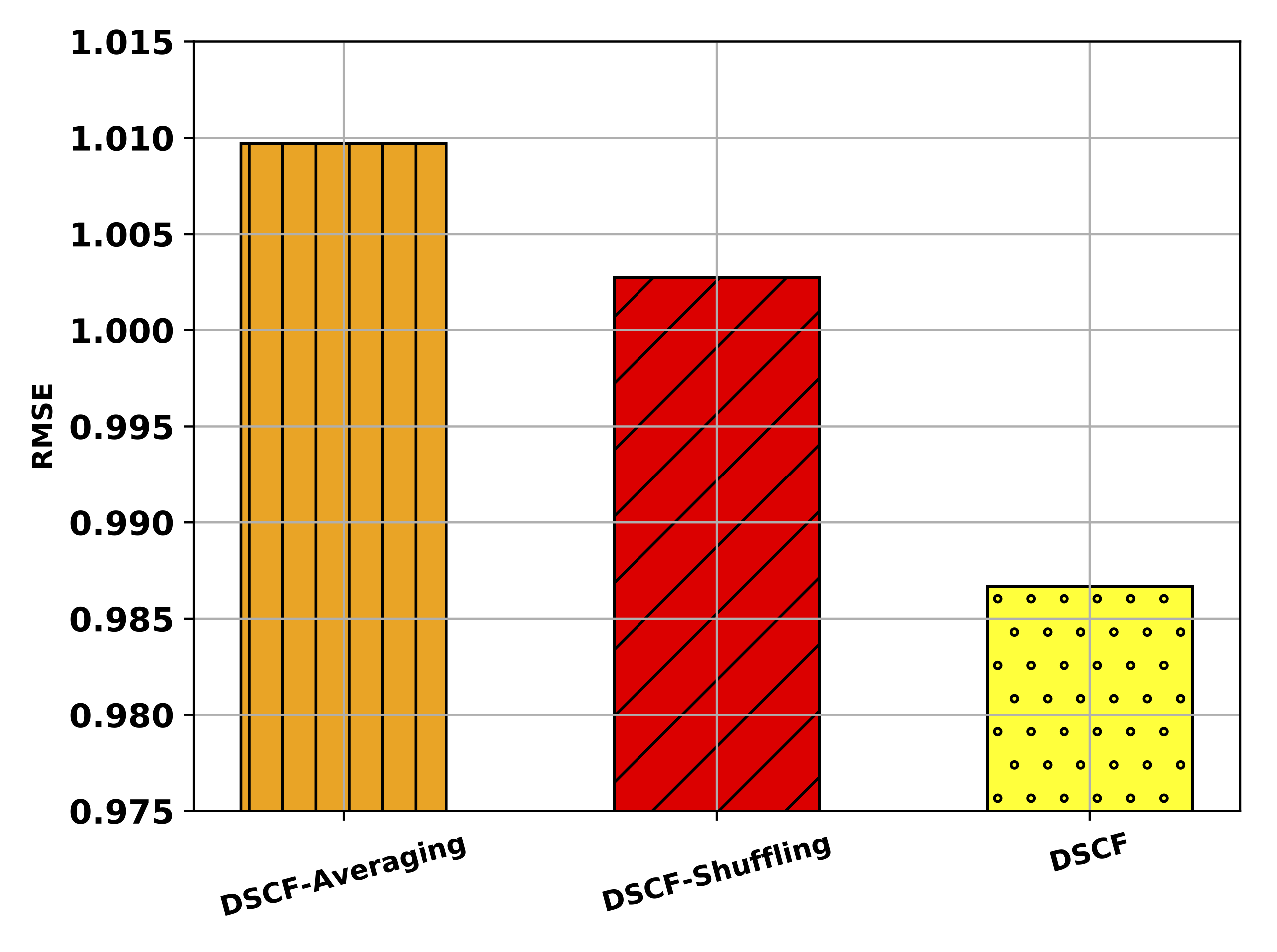}\label{fig:ciao_lstm_rmse}}}
{\subfigure[MAE]
{\includegraphics[width=0.45\linewidth]{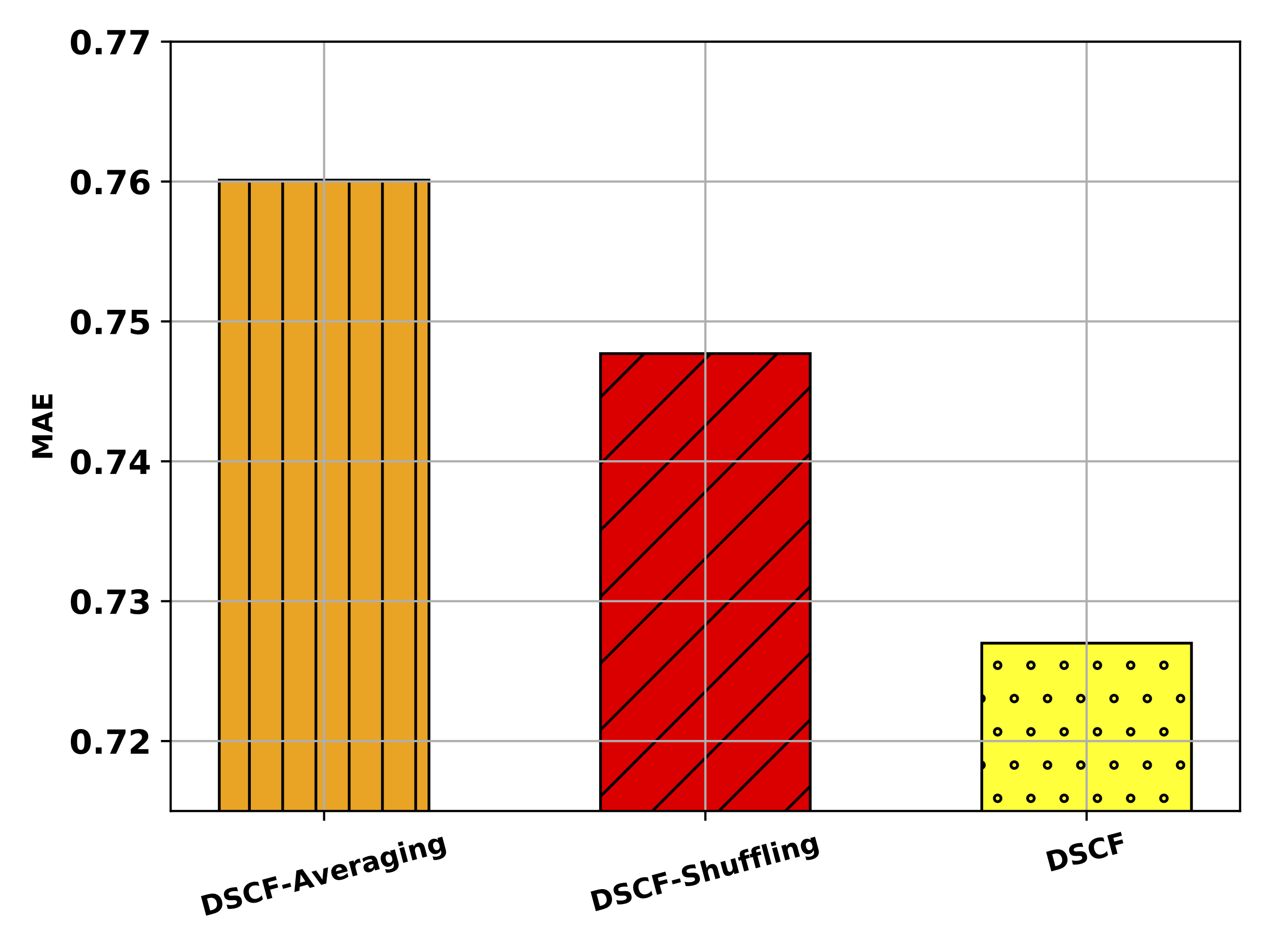}\label{fig:ciao_lstm_mae}}}
\caption{Effect of Bi-LSTM model on Ciao dataset.}\label{fig:ciao_lstm}
\end{figure}
\subsection{Model Component Analysis}

In the previous subsection, we have demonstrated the effectiveness of the proposed framework. To deeply understand DSCF, we compare it with three variants, i.e., DSCF-Opinion, DSCF-Item$\&$Opinion, DSCF-ATT, DSCF-Averaging and DSCF-Shuffling, which are defined as follows:

\begin{itemize}
  \item DSCF-Opinion: This variant uses the item-aware social sequences to represent user's social information; while ignoring the opinions on the user-item interaction.
  \item DSCF-Item$\&$Opinion: Based on DSCF-Opinion, it further eliminates the associated items in the social sequence.
  \item DSCF-ATT: This variant is to study the impact of attention mechanisms on learning ${\bf s}^{u,v}_{(i)}$  and ${\bf s}^{u,v}$. The attention mechanisms $\alpha$ and $\beta$ are removed in this variant.
  \item DSCF-Averaging: This variant replaces Bi-LSTM with averaging the elements in the input of the sequence in the sequence learning layer.
  \item DSCF-Shuffling: This variant randomly shuffles the order of elements in the sequence in the sequence learning layer.
\end{itemize}

The variant DSCF-Averaging considers that all users in the sequence have the same influence to the target user; while DSCF-Shuffling assumes that the influence is not related to  the distance to the target user. These two variants are designed to understand the benefit of adapting Bi-LSTM to capture the item-aware social sequences. 

The results on Ciao are given in Figure~\ref{fig:ciao_uvr} and Figure~\ref{fig:ciao_lstm}. We do not show the results on Epinions since similar observations can be made. From the results, we have the following findings:
\begin{description}
  \item[Item-aware Social Sequences with Opinions. ] We now focus on analyzing the effectiveness of opinions on interactions. From the Figure~\ref{fig:ciao_uvr}, we can see that the performance of DSCF reduces significantly when ignoring the opinions on the user-item interactions in the social sequence (i.e., DSCF-Opinion), which suggests that it is necessary to consider opinions on interactions. In other words, different opinions from a user's friends would affect the user's decision in tremendously different ways.
  \item[Item-aware Social Sequences. ] To recommend a specific item, not all information from users in the sequence is useful; in other words, interactions of these users with related items are more useful. From the results in Figure~\ref{fig:ciao_uvr}, DSCF-Item$\&$Opinion performs worse than DSCF and DSCF-Opinion. These observations support the importance to generate item-aware sequences. In other words, not all information from neighbors are useful for recommending a specific item (e.g., \emph{Spider-man}). Only the information related to this item would be useful (e.g., \emph{Captain America}).
  \item[Attention Mechanisms.] We conducted experiments to verify the effectiveness of the attention mechanism. From the results in Figure~\ref{fig:ciao_uvr}, we can observe that DSCF-ATT obtains worse performance than DSCF. The reason is that not all the user (neighbor)-item interactions in one social sequence contribute equally to learn the representation of item-aware social sequence; and not all these item-aware social sequences have the same importance to the unified representation of all item-aware social sequences. These results demonstrate the benefits of the attention mechanisms on learning ${\bf s}^{u,v}_{(i)}$  and ${\bf s}^{u,v}$.
  \item [Bi-LSTM.] Figure~\ref{fig:ciao_lstm} presents the effect of Bi-LSTM on Ciao dataset. The performance of both DSCF-Averaging and DSCF-Shuffling reduces significantly. It suggests that the Bi-LSTM component is better to learn representations for item-aware social sequences. The reason is that the social sequence reflects the information diffusion to the target user and the influence to the target user should be heterogeneous and related to the distance.
  
\end{description}

\begin{figure}[t]
\centering
{\subfigure[RMSE]
{\includegraphics[width=0.45\linewidth]{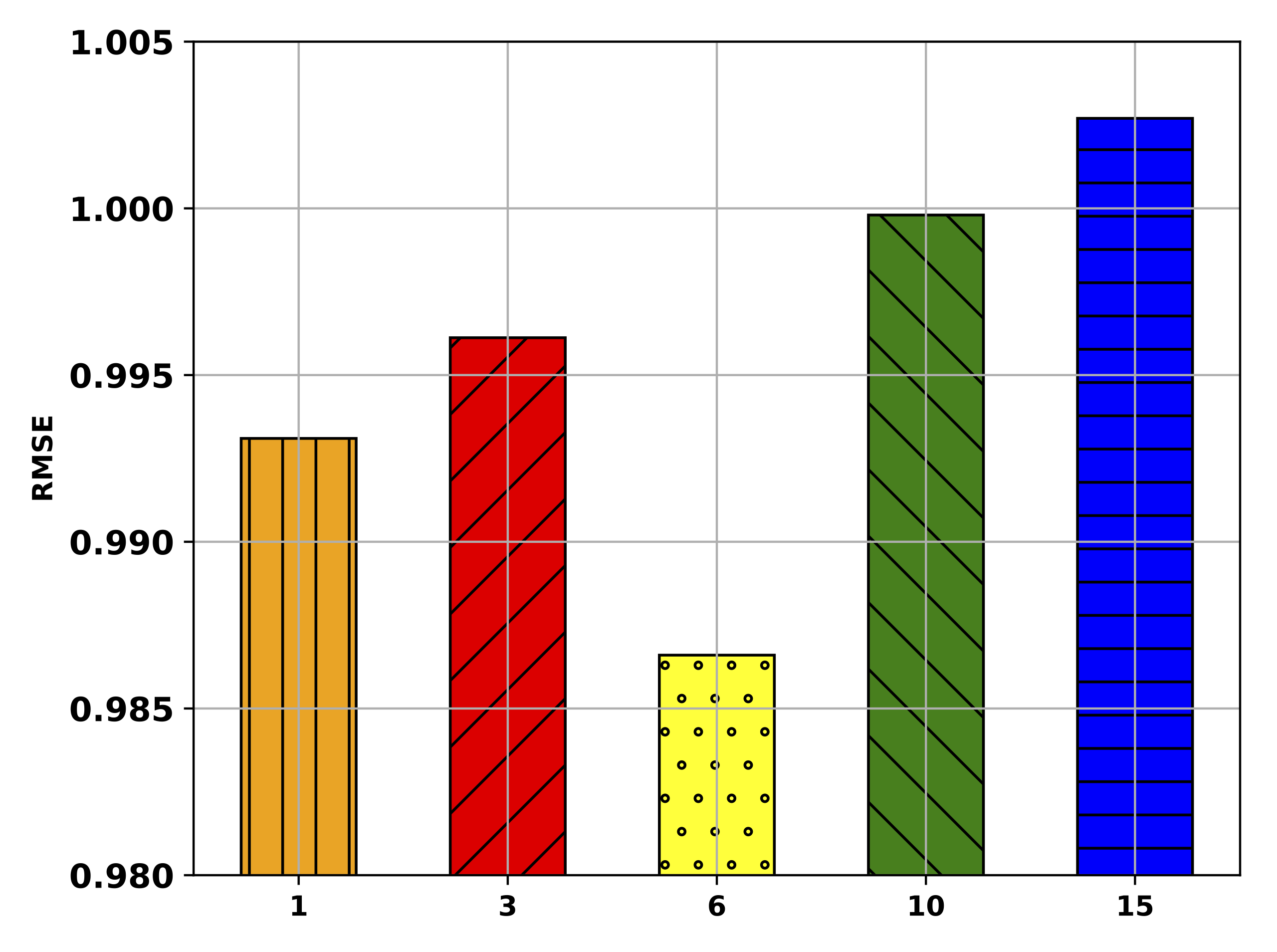}\label{fig:ciao_length_rmse}}}
{\subfigure[MAE]
{\includegraphics[width=0.45\linewidth]{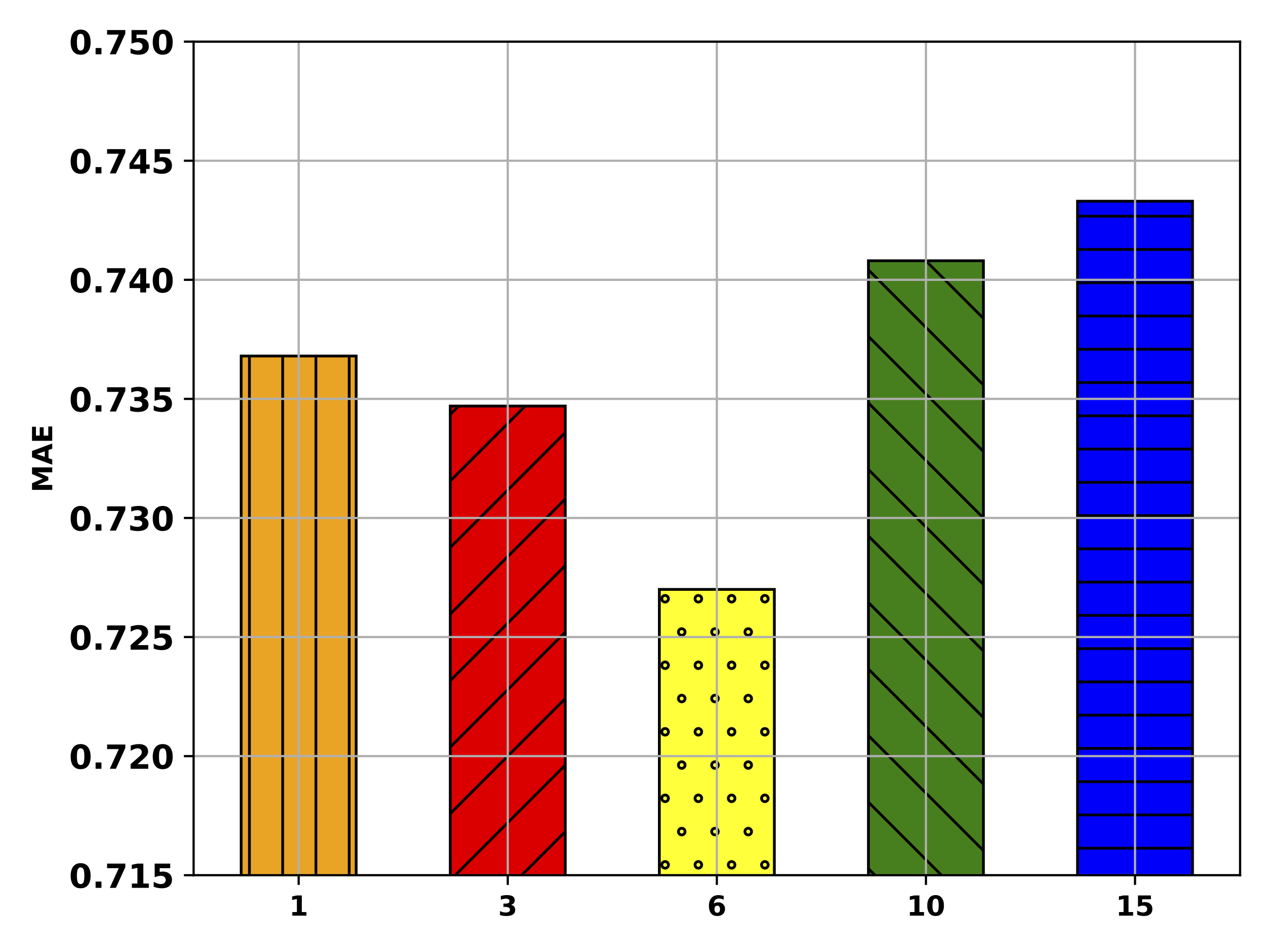}\label{fig:ciao_length_mae}}}
\caption{Performances w.r.t. the length of sequence.}\label{fig:ciao_length}
\end{figure}
\begin{figure}[t]
\centering
{\subfigure[RMSE]
{\includegraphics[width=0.45\linewidth]{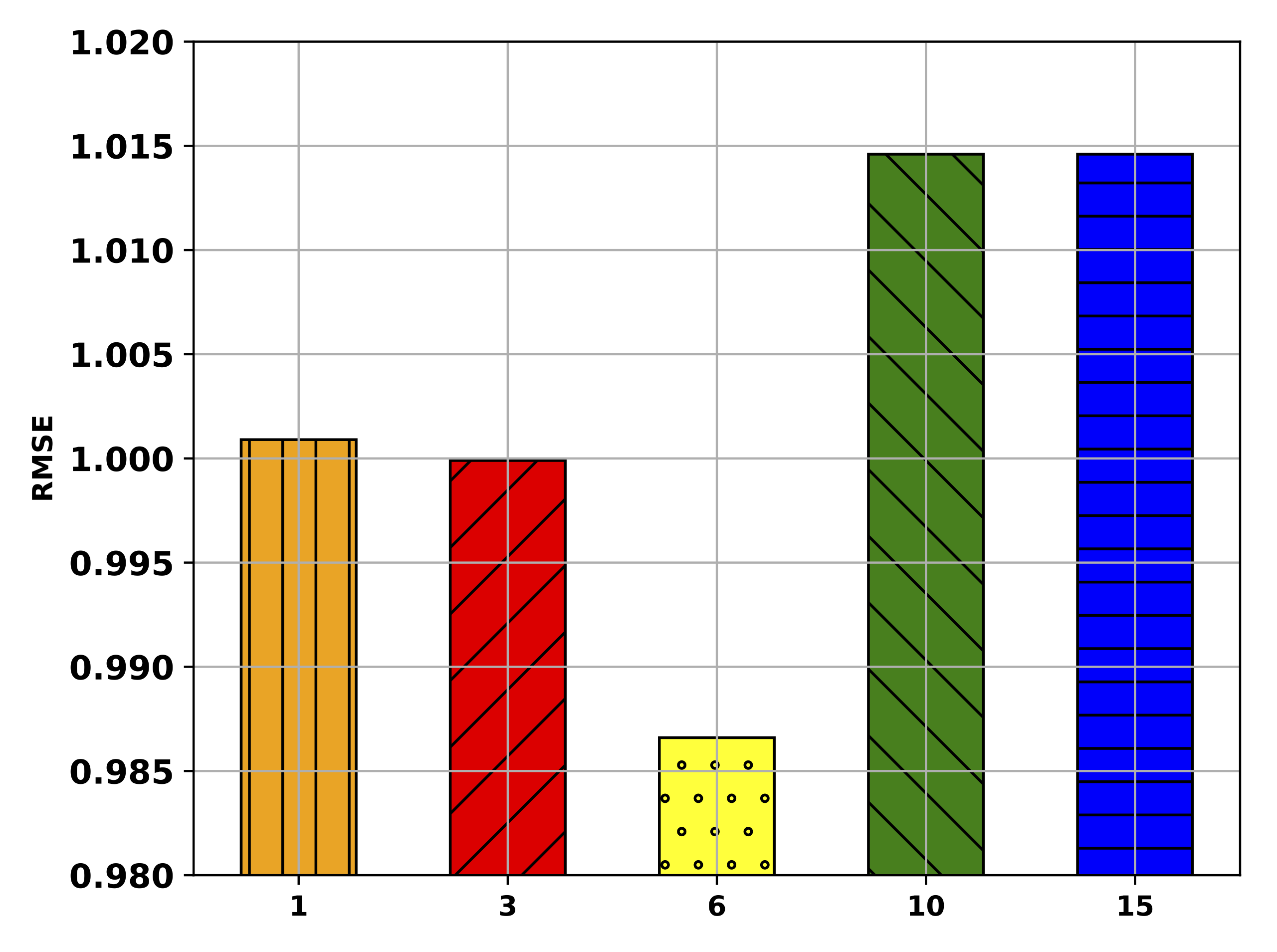}\label{fig:ciao_num_rmse}}}
{\subfigure[MAE]
{\includegraphics[width=0.45\linewidth]{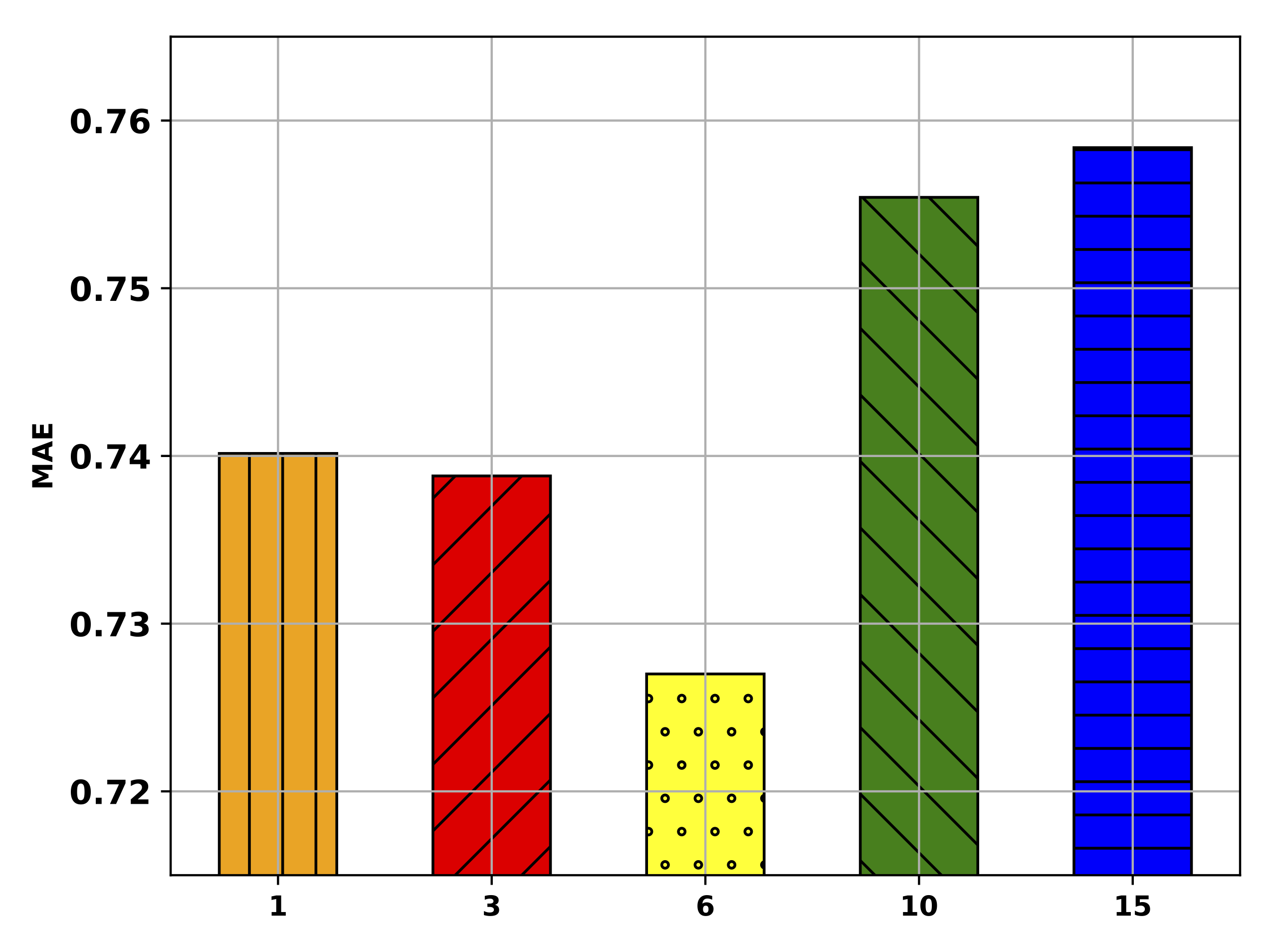}\label{fig:ciao_num_mae}}}
\caption{Performance w.r.t the number of sequences.}\label{fig:ciao_num}
\end{figure}




\subsection{Parameter Analysis}

There are two important parameters of the proposed framework, i.e., the length of each item-aware social sequence and the number of item-aware social sequences. In this subsection, we investigate the impact of these parameters by examining how the performance changes when varying one parameter and fixing others. Similarly, we only show results on Ciao.
\begin{description}
  \item[Effect of the length of sequences $l$. ] Figure~\ref{fig:ciao_length} shows the performance with the varied length of sequences on Ciao. If the length of sequence is one, our model boils down to use the direct neighbors. When the length of sequence increases, the performance tends to increase first. This indicates that the direct neighbors cannot sufficiently capture the useful social information and including distant neighbors could help. However, when the length of sequences becomes too large, the performance degrades as we may introduce too many noises with the distant neighbors. 
  \item[Effect of the number of sequences $H$. ] Figure~\ref{fig:ciao_num} shows how the number of sequences affects the performance of recommendations. Generally more sequences can sufficiently explore the neighborhood of users, which can help us understand social information better; however, it is also risky to generate too many since we may introduce noise as well.
\end{description}

\section{Related Work}
\label{sec:relatedwork}

In this section, we briefly review some researches related to our work. Collaborative filtering~\cite{goldberg1992using}, which captures users' preference towards items utilizing user-item interactions, is the most popular approach to build modern recommender systems. In addition to the user-item interactions, social relations also have potential to help understand users' preference.  Many social recommendation methods~\cite{ma2008sorec,purushotham2012collaborative,wang2016social,tang2013exploiting,tang2016recommendation,zhao2014leveraging,guo2015trustsvd} have shown the effectiveness of including social relations for recommendations. Among them, SoRec~\cite{ma2008sorec} co-factorizes the rating matrix (user-item interaction matrix) and the social relation matrix for recommendation by sharing user latent vectors between them. SoDimRec~\cite{tang2016recommendation} utilizes the heterogeneity of social relations and the weak dependency connections in social networks for recommendation.
A comprehensive survey on social recommendations can be found in ~\cite{tang2013social}.

Recently, deep neural networks have been adopted to enhance recommender systems~\cite{zhu2017next,bai2017neural}. Most of them utilize deep neural networks as feature learning tools to extract features from auxiliary information such as text description of an item~\cite{wang2015collaborative,kim2016convolutional,Chen2018Neural} and visual information of images~\cite{ZhaoLP016}. NeuMF~\cite{He2017NCF}, is a matrix factorization based deep recommendation method, which uses deep neural networks to explore the non-linearity in user-item interactions. NSCR~\cite{wang2017item} extends the NeuMF model by utilizing the social network information as a graph regularization, which enforces nearby neighbors to have similar latent vectors. NSCR addresses the task of cross-domain recommendations for ranking metric, and focuses on how to distill useful signal from an external social network (e.g., Facebook and Twitter) on the cross-domain task, while our model focuses on how to learn the social information from the user-user interaction in the same e-commerce platform, rather than external social network. ARSE~\cite{sun2018attentive} proposes the problem of temporal social recommendation for ranking metric, which has dynamic and static part to model the dynamic and static preferences of users. ARSE targets on the dynamic preferences of the recommendation, rather that the social information.

Most related to our task with neural networks includes DLMF ~\cite{deng2017deep}, GCMC~\cite{berg2017graph}, DeepSoR~\cite{DeepSoR2018}, GraphRec~\cite{fan2019graph} and DASO~\cite{fan2019deep}.
DeepSoR~\cite{DeepSoR2018} first represents users using pre-trained node embedding technique, and further utilizes deep neural networks to capture non-linear features in social relations and integrate them into probabilistic matrix factorization. DASO~\cite{fan2019deep} proposes a deep adversarial social recommendation framework, which adopts a bidirectional mapping method to transfer users' information between social domain and item domain using adversarial learning. GraphRec~\cite{fan2019graph} harness the power of graph neural networks (GNNs) techniques to model graph data in social recommendations by aggregating the both user-item interactions information and direct social neighbors. However, these deep social recommendation methods cannot take full advantages of social networks. In this paper, we propose a deep social recommendation framework which can sufficiently exploit the social network information for recommendations.

\section{Conclusion and Future work}
\label{sec:conclusion}

We have presented a \textbf{D}eep \textbf{S}ocial \textbf{C}ollaborative \textbf{F}iltering (\textbf{DSCF}) which can exploit the social information with various aspects for recommendations. Particularly, we propose to utilize the random walk to generate item-aware social sequences, which consider information from not only direct neighbors but also distant neighbors. In addition, we also introduce a novel way to capture neighbors' opinions when modeling user-item interactions. Finally, the Bi-LSTM with attention mechanism is proposed to extract feature for the social sequence. Our experiments reveal that the item-aware sequences and the opinion information play a crucial role in modeling social information.  Comprehensive experiments on two real-world datasets show the effectiveness of our model. In this work, we only utilize the user-item interactions to measure the similarity between items, while rich side information may be associated with items, such as the textual description, and the visual content of images. Therefore, incorporating side information would be considered as an interesting future direction.

\begin{acks}

The work is partly supported by NSFC-Guangdong Joint Fund under project U1501254, Science Technology and Innovation Committee of Shenzhen Municipality Under project JCYJ20170818095109386, and an internal research grant (project no. 9B0V) from the Hong Kong Polytechnic University. Yao Ma and Jiliang Tang are supported by the National Science Foundation (NSF) under grant numbers IIS-1714741, IIS-1715940, IIS-1845081 and CNS-1815636, and a grant from Criteo Faculty Research Award.

\end{acks}

\bibliographystyle{ACM-Reference-Format}
\balance
\bibliography{references/references}
\end{document}